\documentclass[technote,9pt]{./IEEEtran}

\ifCLASSINFOpdf
\else
\fi
\usepackage{graphicx}
\DeclareGraphicsExtensions{.eps}

\usepackage[cmex10]{amsmath}
\newtheorem{theorem}{Theorem}[section]

\newtheorem{lemma}[theorem]{Lemma}

\usepackage{array}

\usepackage[tight,footnotesize]{subfigure}
\usepackage{url}

\begin{document}

%
\title{Twenty Questions Games Always End With Yes}
%
%
%

\author{John~T.~Gill~III,~\IEEEmembership{Member,~IEEE,}
        and~William~Wu
\thanks{John T. Gill III and William Wu are with the Department
of Electrical Engineering, Stanford University, Stanford,
CA, 94305 USA.}
}

\maketitle

\begin{abstract}
Huffman coding is often presented as the optimal solution to Twenty Questions. However, a caveat is that Twenty Questions games always end with a reply of ``Yes," whereas Huffman codewords need not obey this constraint. We bring resolution to this issue, and prove that the average number of questions still lies between $H(X)$ and $H(X)+1$.
\end{abstract}

\begin{IEEEkeywords}
Huffman coding, entropy, twenty questions game, Gallager's redundancy bound
\end{IEEEkeywords}

%
\IEEEpeerreviewmaketitle

\section{Introduction}
%
%
%
%
Twenty Questions  is a classic parlour game involving an answerer and a questioner. The questioner must guess what object the answerer is thinking of, but is only allowed to ask questions whose answers are either ``Yes" or ``No". 
Popular initial questions include: ``Is it an animal? Is it a vegetable? Is it a mineral?" The name of the game arises from the fact that if one bit of information could be acquired from each question, then twenty questions can distinguish between $2^{20}$ different objects, which should be more than sufficient. 

Courses in information theory often cast Huffman coding as the optimal approach to Twenty Questions. Given the set of possible objects and their probabilities, the questioner associates a Huffman codeword with each object,  and then inquires about each bit of the codeword that the questioner is thinking of. The average number of questions is the Huffman tree's average depth, which is no less than $H(X)$, and less than $H(X)+1$, where $X$ is the random variable indicating which of $n$ objects the answerer is thinking of.

However, upon further thought, there is a disparity between Huffman coding and how Twenty Questions games are played. Namely, real-world Twenty Questions games always terminate with the questioner pinpointing a specific object (e.g., ``Is it a {\it tank}?" \cite{billandted}), to which the answerer replies, ``Yes!" In terms of source coding, this is equivalent to enforcing what we call the {\it terminating yes constraint}:  all codewords must terminate with ``1". Yet Huffman codes do not satisfy this constraint! In short, Huffman trees determine $X$, but do not specify $X$.
 
In this paper, we first provide an example showing that simply appending branches to a Huffman tree may not produce the optimal Twenty Questions tree. We then prove that even under the terminating yes constraint, the average number of questions lies strictly between $H(X)$ and $H(X) + 1$.

\section{Bar Bet: Guessing One of Four Objects}

Since Huffman coding solves Twenty Questions without a terminating yes, a natural idea is to first produce the Huffman tree, and then append branches to it so the terminating yes
 constraint is  satisfied. Call the result an {\it augmented Huffman tree}. In the following example, we show that augmented Huffman trees may not be optimal Twenty Questions trees.

\begin{figure}[ht]
\centering
\subfigure[]{
\includegraphics[scale=0.55]{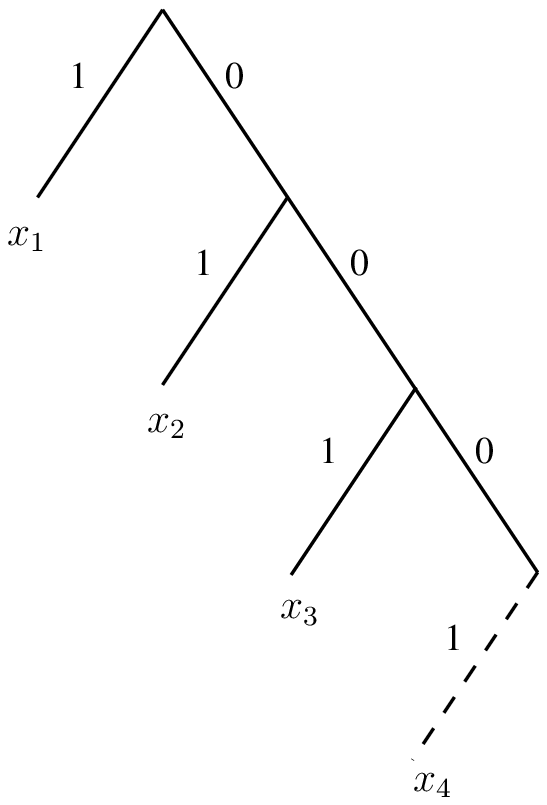}
\label{fig:4objectsUnary}
}
\qquad
\subfigure[]{
\includegraphics[scale=0.55]{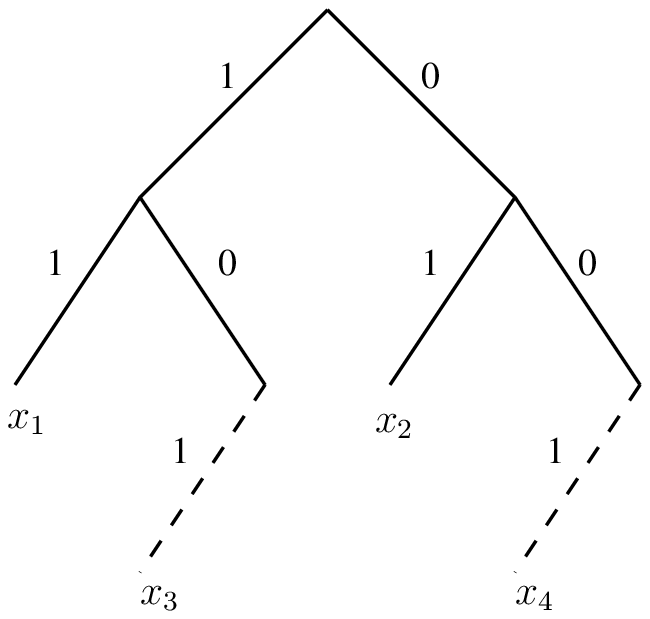}
\label{fig:4objectsBalanced}
}
\caption[Four leaf trees.]{Four leaf trees. \subref{fig:4objectsUnary} Unary code. 
\label{fig:4objects}
\subref{fig:4objectsBalanced} Balanced code.}
\end{figure}

Suppose there are only four objects the answerer could be thinking of. Denote them by $x_1,x_2,x_3,x_4$, with corresponding probabilities $p_1 \geq p_2 \geq p_3 \geq p_4$. Figure \ref{fig:4objects} shows the only two four-leaf questioning trees possible up to graph isomorphism, where the dashed edges have been added to accommodate the terminating yes
 constraint. Although there are many possible assignments of objects to leaves, the assignments shown in Figure \ref{fig:4objects} are the only reasonable candidates which place lower probability objects at shallower depths.


One naturally imagines that the choice of a questioning tree should depend on the probability distribution. For instance, if the probabilities are close to uniform, we would guess that the balanced tree is better.  However, if we let $Q_1$ and $Q_2$ denote the average number of questions used by the unary and balanced trees, respectively, then
\begin{align*}
Q_1 
&= p_1 + 2 p_2 + 3 p_3 + 3 p_4 = 1 + p_2 + 2 p_3 + 3 p_4, \\
Q_2 
&= 2(p_1 + p_2) + 3 p_3 + 3 p_4 = 2 + p_3 + p_4,
\end{align*}
and the difference is
\begin{align*}
Q_2 - Q_1
&= 2 + p_3 + p_4 - 1 - p_2 - 2p_3 - 3p_4 \\
&= 1 - (p_2 + p_3 + 2p_4) \\
&= p_1 - p_4 \geq 0,
\end{align*}
with equality if and only if the distribution is uniform. Apparently the unary tree dominates the balanced tree, regardless of the probabilities! We think this makes for a good bar bet. 

This example demonstrates that augmenting a Huffman tree does not necessarily produce the optimal Twenty Questions tree. For example, if the probabilities were $(3/10,3/10,2/10,2/10)$, then the resulting augmented Huffman tree would yield the balanced tree, although the unary tree is better. In fact, among all distributions for which the Huffman algorithm produces a balanced tree, the maximum difference in the average number of questions required by the balanced and unary trees approaches 1/3, and is achieved with the distribution $(\frac{1}{3} - \epsilon,\frac{1}{3} - \epsilon,\frac{1}{3} - \epsilon,3\epsilon)$.

\section{Entropy Bounds On The Average Number of Questions}

Let $L_H$ be the average depth of the Huffman tree, and let  $L_{yes}
$ be the average depth of the optimal Twenty Questions tree. In this section, we prove
\[
H(X) < L_{yes}
 < H(X)+1.
\]
Note that these are the same bounds satisfied by $L_H$, except for the strict inequality in the lower bound. We first require two Lemmas.

\begin{lemma}[Half-Bit Lemma]
\label{lem:halfBit}
A binary tree that does not satisfy the terminating yes constraint can be modified to satisfy it while adding no more than 1/2 to the average depth.  
\end{lemma}
\begin{IEEEproof}
Let $T$ be a tree that does not satisfy the terminating yes constraint. By appending a branch to all leaves whose codewords end with 0, we can construct an augmented tree $T'$ that does satisfy it. (This forces all leaves to sway in the same direction.) To minimize the increase in average depth, interchange siblings in $T$ as necessary so that the lower probability sibling is always the one that receives the appended branch. Consequently, if the average length of $T$ is $L$, the average length of $T'$ will be no more than $L + 1/2$. 
\end{IEEEproof}

\begin{lemma}[Gallager's Redundancy Bound]
\label{lem:gallager1978variations}
For all finite distributions, $L_H - H(X) \leq p_1 + \sigma$, where $p_1$ is the largest probability, and $\sigma := 1 - \log_2 e + \log_2 (\log_2 e) \approx 0.086$.
\end{lemma}
\begin{IEEEproof}
See Gallager \cite{gallager1978variations}.
\end{IEEEproof}

\begin{theorem}
\label{thm:entropyBounds}
$H(X) < L_{yes}
 < H(X)+1$.
\end{theorem}
\begin{IEEEproof}
We first establish the lower bound. By pruning the appended branches of the optimal Twenty Questions tree, we have a new tree of reduced average depth in which every internal node has two children. Amongst all such trees, the Huffman tree has lowest average depth, so $L_H < L_{yes}
$. Lastly, $H(X) \leq L_H$ (see Cover and Thomas \cite{cover2006elements}). 

For the upper bound, we consider two cases. First, suppose $p_1 < 0.4$. From Lemma \ref{lem:gallager1978variations}, 
\begin{align*}
L_H - H(X) &\leq  p_1 + \sigma  <  1/2\,.
\end{align*}
Adding 1/2 to both sides and rearranging, 
\[
L_H + 1/2 < H(X) + 1\,.
\]
From Lemma \ref{lem:halfBit}, $L_{yes}
 \leq L_H + 1/2$. Thus,
\[
L_{yes}
 < H(X) + 1\,.
\] 

When $p_1 \geq 0.4$, we prove the upper bound by induction on the number of objects. Let $X$ be a random variable taking $n$ possible values, and let $\hat{T}$ be the tree with minimum average depth under both the terminating yes constraint {\it and} the additional constraint that the most probable object has a codeword of length one.
 This tree $\hat{T}$ is illustrated in Figure \ref{fig:inductionArgumentTree}. While this additional constraint may result in a suboptimal tree,  we will show that $\hat{T}$ satisfies the desired upper bound regardless, and thus the optimal Twenty Questions tree does also. 

\begin{figure}[htbp]
\begin{center}
\includegraphics[scale=0.7]{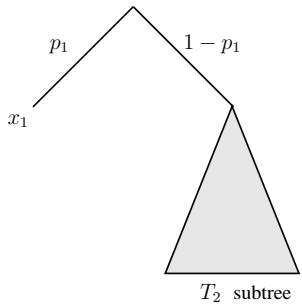}
\caption{Tree $\hat{T}$ used in the induction argument when $p_1 \geq 0.4$.}
\label{fig:inductionArgumentTree}
\end{center}
\end{figure}

Let $\hat{L}$ denote the average depth of $\hat{T}$, and let $T_2$ denote the right subtree containing $n-1$ leaves. Then 
\begin{align*}
\hat{L} &= 1 + (1-p_1)  L(T_2)
\end{align*}
where $L(T_2)$ denotes the average depth of $T_2$. Also, by the grouping law for entropy, 
\begin{align*}
H(X) &= H(p_1) + (1-p_1) H(X_2)
\end{align*}
where $X_2$ is the random variable given by the remaining $n-1$  normalized probabilities $\left(\frac{p_2}{1-p_1},\frac{p_3}{1-p_1},\ldots,\frac{p_n}{1-p_1}\right)$. Subtracting these equations, 
\[
\hat{L} - H(X)  
= 1 - H(p_1) + (1-p_1) (L(T_2) - H(X_2))
\]
By construction of $\hat{T}$, it follows that $T_2$ is an optimal Twenty Questions
 tree for $X_2$. By the induction hypothesis, $L(T_2) - H(X_2) < 1$, and thus
\[
\hat{L} - H(X)  
\leq 
2 - (H(p_1) + p_1).
\]
Since $p_1$ could be any value in $[0.4,1]$, we want the largest upper bound, to cover all our bases. Setting $p_1 = 1$,
\[
\hat{L} - H(X)  
\leq 
1.
\]
Thus, 
\[
L_{yes}
 \leq \hat{L} \leq H(X) + 1.
\]
\end{IEEEproof}

Lastly, by comparing the bounds
\begin{align}
\label{eq:classicalBounds}
H(X) &\leq L_H < H(X) + 1 \\
 H(X) &< L_{yes} < H(X) + 1 
 \end{align}
we conclude that
\begin{equation}
H(X) \leq  L_H < L_{yes} < H(X) + 1.
\end{equation}
Since the classical bounds in Equation \ref{eq:classicalBounds} are tight, it follows that the bounds in Theorem \ref{thm:entropyBounds} are also tight.

\section{Conclusion}

We have provided resolution to a disconnect between the Twenty Questions game and Huffman coding. Although Twenty Questions games always end with ``Yes", thankfully the average number of questions they require is still within one of the entropy -- a nice answer to a simple problem. As Forrest Gump would say, ``One less thing to worry about."

\section*{Acknowledgment}

The authors graciously  thank Thomas Cover for conceiving this problem and sharing many enlightening discussions pertaining to it, as well as Paul Cuff for his expertise in the most excellent adventures of Bill and Ted.

\ifCLASSOPTIONcaptionsoff
  \newpage
\fi



%

%


\begin{IEEEbiographynophoto}{John T. Gill III}
received the B.S. in Applied Mathematics from Georgia Tech in 1967, and the M.A. and Ph.D. degrees in Mathematics from the University of California Berkeley, in 1969 and 1972, respectively. Since then he has been Associate Professor of Electrical Engineering at Stanford University, Stanford, CA. His research interests include computational complexity theory, information theory, probabilistic computation, and efficient representations of data. His investigations of the P vs. NP question include the Baker-Gill-Solovay relativization theorem.
\end{IEEEbiographynophoto}


\begin{IEEEbiographynophoto}{William Wu}
received the B.S. degree in Electrical Engineering and Computer Science from the University of California, Berkeley in 2003. At Stanford University, he received M.S. degrees in Electrical Engineering and Mathematics in 2005 and 2009, and recently defended his Ph.D. in Electrical Engineering. He is the creator of {\tt wuriddles.com}, a large archive of mathematical puzzles. His research interests include sampling and reconstruction, signal processing, communications, and recreational math. 
\end{IEEEbiographynophoto}




\end{document}